# The "Out-Longitudinal" Cross Term and Other Model Independent Features of the Two-Particle HBT Correlation Function

S. Chapman[a], P. Scotto[b], and U. Heinz[b]

[a]Los Alamos National Laboratory, Los Alamos, NM 87545, USA

[b]Institut für Theoretische Physik, Universität Regensburg, D-93040 Regensburg, Germany

Using two specific models and a model independent formalism, we show that an "out-longitudinal" cross term should be included in any gaussian fits to correlation data. In addition, we show that correlation radii (including the cross term) measure lengths of homogeneity within the source, not necessarily geometric sizes.

## 1. Introduction

Experimentally measured Hanbury-Brown Twiss (HBT) correlations between two identical particles are typically fit by gaussians of the form [1,2]

$$C(\mathbf{q}, \mathbf{K}) = 1 \pm \lambda \exp\left[-q_s^2 R_s^2(\mathbf{K}) - q_\perp^2 R_\perp^2(\mathbf{K}) - q_L^2 R_L^2(\mathbf{K})\right] , \qquad (1)$$

where $\mathbf{q} = \mathbf{p}_1 - \mathbf{p}_2$, $\mathbf{K} = \frac{1}{2}(\mathbf{p}_1 + \mathbf{p}_2)$, the + (−) sign is for bosons (fermions), and the coordinate system is defined as follows [3]: The "longitudinal" or $\hat{z}$ (subscript $L$) direction is parallel to the beam; the "out" or $\hat{x}$ (subscript $\perp$) direction is parallel to the component of $\mathbf{K}$ which is perpendicular to the beam; and the "side" or $\hat{y}$ (subscript $s$) direction is the remaining direction. We claim that significantly more can be learned and better fits achieved if an "out-longitudinal" cross term is included in the following way [4]

$$C(\mathbf{q}, \mathbf{K}) = 1 \pm \lambda \exp\left[-q_s^2 R_s^2(\mathbf{K}) - q_\perp^2 R_\perp^2(\mathbf{K}) - q_L^2 R_L^2(\mathbf{K}) - 2q_\perp q_L R_{\perp L}^2(\mathbf{K})\right] , \qquad (2)$$

where $R_{\perp L}^2$ is a parameter which can be either positive or negative.

To see how this cross term arises in two-particle correlations, we use the following well established theoretical approximation [5,6]

$$C(\mathbf{q}, \mathbf{K}) \simeq 1 \pm \frac{|\int d^4x\, S(x, K)\, e^{iq \cdot x}|^2}{|\int d^4x\, S(x, K)|^2} , \qquad (3)$$

where $q_0 = E_1 - E_2$ and $K_0 = E_K = \sqrt{m^2 + |\mathbf{K}|^2}$. Here $S(x, K)$ is a function which describes the phase space density of the emitting source. For pairs with $|\mathbf{q}| \ll E_K$, we can use the approximation $q_0 \simeq \beta_\perp q_\perp + \beta_L q_L$, where $\beta_i = K_i/E_K$.

## 2. Stationary Gaussain Sources

Consider the following azimuthally symmetric gaussian emission function:

$$S(x,K) = f(K)\exp\left[-\frac{x^2+y^2}{2R^2} - \frac{z^2}{2L^2} - \frac{(t-t_0)^2}{2(\delta t)^2}\right]. \tag{4}$$

Using (3), one can see that the corresponding correlation function takes the form

$$C(\mathbf{q},\mathbf{K}) = 1 \pm \exp[-q_s^2 R^2 - q_\perp^2(R^2+\beta_\perp^2(\delta t)^2) - q_L^2(L^2+\beta_L^2(\delta t)^2) - 2q_\perp q_L \beta_\perp \beta_L (\delta t)^2], \tag{5}$$

so the $q_\perp q_L$ cross term provides a measurement of the duration of particle emission ($\delta t$).

The effect of the cross term is easiest to see by looking at the "out-longitudinal" projection of the correlation function. Figures 1 and 2 show contour plots of the "out-longitudinal" correlator for a source featuring $R = L = c\delta t = 3$ fm and pairs characterized by $q_s = 0$, $\beta_\perp = \sqrt{.5}\,c$. The outer contours are for $|C - 1| = 0.1$, and each successively smaller contour represents an increase of 0.1. In figure 1 $\beta_L = 0$, so the cross term vanishes, thus causing the major and minor axes of the elliptical contours to coincide with the "out" and "longitudinal" axes. In figure 2, however, $\beta_L = \sqrt{.5}\,c$, and the nonvanishing cross term causes a rotation in the major and minor axes. This effect should be easily observable, and has in fact already been measured in NA35 correlation data [7].

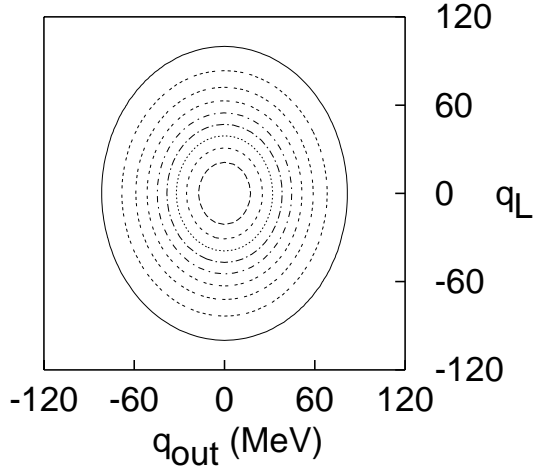 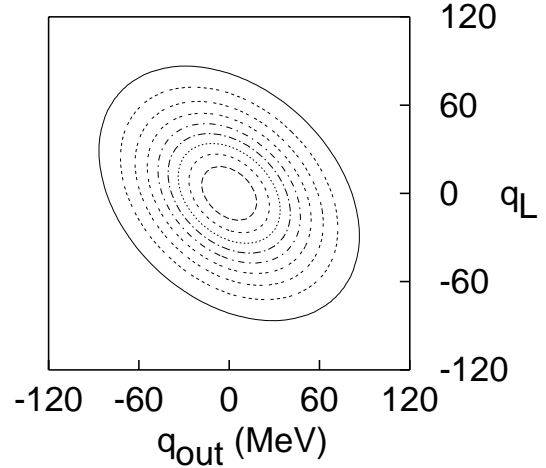

Figure 1: Eq. (5) with $\beta_L = 0$.      Figure 2: Eq. (5) with $\beta_L = \sqrt{1/2}\,c$.

## 3. Model Featuring Boost-Invariant Expansion

Now we turn to a more realistic model, similar to those in [8]. In the center of mass frame of an expanding fireball, we define the following emission function [4,6]:

$$S(x,K) = \frac{\tau_0 m_t \text{ch}(\eta-Y)}{(2\pi)^3 \tau \sqrt{2\pi(\delta\tau)^2}} \exp\left[-\frac{K\cdot u(x)}{T} - \frac{x^2+y^2}{2R_G^2} - \frac{\eta^2}{2(\delta\eta)^2} - \frac{(\tau-\tau_0)^2}{2(\delta\tau)^2}\right], \tag{6}$$

where $T$ is a constant freeze-out temperature, $\tau = \sqrt{t^2-z^2}$, $\eta = \frac{1}{2}\ln[(t+z)/(t-z)]$, $m_t = \sqrt{m^2+K_\perp^2}$, and $Y$ is the rapidity of a particle with momentum $\mathbf{K}$. We consider a flow which is non-relativistic transversally but which exhibits Bjorken expansion longitudinally,

$$u(x) \simeq \left(\left(1+\tfrac{1}{2}(v\rho/R_G)^2\right)\text{ch}\eta,\ (vx/R_G),\ (vy/R_G),\ \left(1+\tfrac{1}{2}(v\rho/R_G)^2\right)\text{sh}\eta\right), \tag{7}$$

where $\rho = \sqrt{x^2 + y^2}$ and $v \ll 1$ is the transverse flow velocity of the fluid at $\rho = R_G$.

To generate an "out-longitudinal" correlation function, we plug in some numbers and then numerically integrate eq. (3), using (6) and (7). For simplicity, we consider a pion source with no transverse flow ($v = 0$) which freezes out instantaneously ($\delta\tau = 0$) with the following other source parameters: $R_G = 3$ fm, $\tau_0 = 4$ fm/c, $\delta\eta = 1.5$, and $T = 150$ MeV. In [6] we show that when the emission function (6) with these parameters is integrated over spacetime, it produces a very reasonable one-particle distribution. Figure 3 shows the "out-longitudinal" correlator for pairs with $Y = -2$, $K_\perp = 200$ MeV, and $q_s = 0$. Figure 4 shows the same correlation calculated in the LCMS frame, which is the longitudinally boosted frame defined by $\beta_L = 0$ [2]. Since both figures feature rotations of the major and minor axes, it is apparent that the correlations both in the fixed and LCMS frames would be much better fit by eq. (2) than by eq. (1).

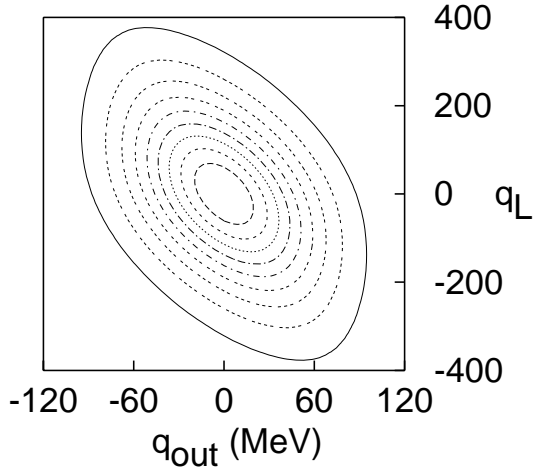 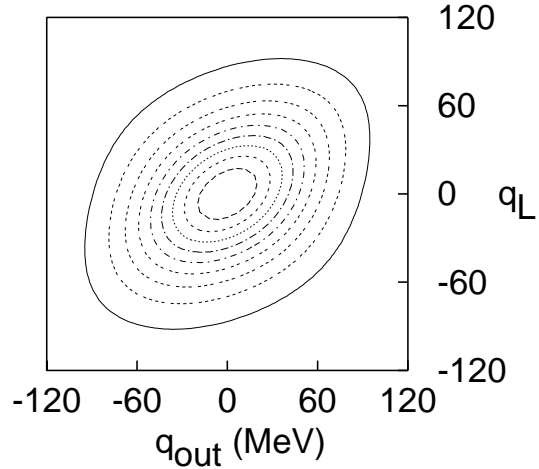

Figure 3: Correlation calculated numerically using (3), (6) and (7) in the fixed center of mass frame.

Figure 4: Same as fig.3, but calculated in the LCMS frame.

## 4. Model Independent Derivation of the Correlation Function

By making a saddle point approximation to a general emission function, we can derive model independent expressions for all of the correlation radii [4,6]. We define the spacetime saddle point $\bar{x}$ of the emission function $S(x, K)$ through the four equations

$$\frac{d}{dx_\mu} \ln S(x, K)\Big|_{\bar{x}} = 0 \qquad (8)$$

where $\mu = \{0, 1, 2, 3\}$. Essentially the saddle point is that point in spacetime which has the maximum probability of emitting a particle with momentum $\vec{K}$. A saddle point approximation for $S(x, K)$ can then be made in the following way [6]

$$S(x, K) \simeq S(\bar{x}, K) \exp\left[-\sum_\mu \frac{(x_\mu - \bar{x}_\mu)^2}{2\lambda_\mu^2} - \sum_{\mu > \nu} B_{\mu\nu}(x_\mu - \bar{x}_\mu)(x_\nu - \bar{x}_\nu)\right], \qquad (9)$$

where we define lengths of homogeneity and source mixing parameters by

$$\lambda_\mu(\vec{K}) = \left[-\frac{d^2}{dx_\mu^2} \ln S(x, K)\Big|_{\bar{x}}\right]^{-1/2} \quad \text{and} \quad B_{\mu\nu}(\vec{K}) = -\frac{d}{dx_\mu}\frac{d}{dx_\nu} \ln S(x, K)\Big|_{\bar{x}}. \qquad (10)$$

4Given the general gaussian approximation of (9), the correlation radii take the following model independent form for azimuthally symmetric systems [4,6]

$$R_\perp^2 = \left\langle (x - \beta_\perp t)^2 \right\rangle - \left\langle x - \beta_\perp t \right\rangle^2, \qquad R_L^2 = \left\langle (z - \beta_L t)^2 \right\rangle - \left\langle z - \beta_L t \right\rangle^2,$$
$$R_s^2 = \left\langle y^2 \right\rangle, \qquad R_{\perp L}^2 = \left\langle (x - \beta_\perp t)(z - \beta_L t) \right\rangle - \left\langle x - \beta_\perp t \right\rangle \left\langle z - \beta_L t \right\rangle, \qquad (11)$$

where we use the notation

$$\left\langle \xi \right\rangle \equiv \left\langle \xi \right\rangle(\mathbf{K}) = \frac{\int d^4x\, \xi\, S(x, K)}{\int d^4x\, S(x, K)}. \qquad (12)$$

LCMS radii can be found by setting $\beta_L = 0$ in (11) and using in eq. (12) a form of $S(x, K)$ which is valid in the LCMS frame. We can see that in this frame, $R_{\perp L}^2$ provides information about any $z$-$t$ and/or $z$-$x$ correlations of the source (as seen in the LCMS). Obviously, for the model shown in figure 4, these correlations are not insignificant.

It has been shown that when the $B_{\mu\nu}$ mixing parameters are small, the correlation radii simply measure the lengths of homogeneity $\lambda_\mu$ of the source [9]. For example, for the source defined by eqs. (6) and (7), the "side" radius takes the form [4,6]

$$R_s = \lambda_2 = R_G(1 + m_t v^2/T)^{-1/2}. \qquad (13)$$

Here we can see explicitly that when transverse flow is present ($v \neq 0$), the length of homogeneity measured by the "side" radius is smaller than the geometrical transverse radius. Due to the $m_t$ dependence, this reduction effect is more pronounced for higher momentum and/or larger mass particles.

Using two specific model sources and a model independent formalism, we have shown that there is no a priori reason why an $R_{\perp L}^2$ cross term should be excluded from gaussian fits to experimental correlation data. Not only will the new parameter reveal more information about the source, its inclusion will undoubtedly increase the accuracy of the other fitted radii.

## REFERENCES

1. NA35 Coll., G. Roland et al., Nucl. Phys. **A566** (1994) 527c;
   NA44 Coll., M. Sarabura et al., Nucl. Phys. **A544** (1992) 125c;
   E802 Coll., T. Abbott et al., Phys. Rev. Lett. **69** (1992) 1030.
2. NA35 Coll., D. Ferenc et al., Nucl. Phys. **A544** (1992) 531c;
   NA44 Coll., H. Beker et al., Z. Phys. **C64** (1994) 209.
3. G. Bertsch, M. Gong, and M. Tohyama, Phys. Rev. C **37** (1988) 1896.
4. S. Chapman, P. Scotto and U. Heinz, Regensburg preprint TPR-94-28,
   hep-ph/9408207, submitted to Phys. Rev. Lett.
5. S. Pratt, T. Csörgő and J. Zimanyi, Phys. Rev. **C42** (1990) 2646.
6. S. Chapman, P. Scotto and U. Heinz, Regensburg preprint TPR-94-29,
   hep-ph/9409349, submitted to Heavy Ion Physics (Acta Phys. Hung., New Series).
7. NA35 Coll., T. Alber et al., talk at Quark Matter '95, these proceedings.
8. T. Csörgő, Lund U. preprint LUNFD6 (NFFL-7081) (1994), Phys. Lett. B, in press;
   T. Csörgő and B. Lørstad, Lund U. preprint LUNFD6 (NFFL-7082) (1994).
9. Yu. Sinyukov, Talk presented at Nato Advanced Research Workshop "Hot Hadronic Matter: Theory and Experiment", Divonne June 27 - July 1, 1994.